\documentclass[12pt]{article}
\usepackage{cite,epsfig,amssymb,amsmath,graphicx,color}
\newcommand{\nn}{\nonumber}

\newcommand{\tcr}{\textcolor{red}}

\newcommand{\1}{1\kern -3pt \mathrm{l}}

\begin{document}

\begin{center}
{{{\Large \bf Super Yang-Mills Theories with  Inhomogeneous Mass Deformations}
}\\[17mm]
~~Yoonbai Kim$^{1}$,~~O-Kab Kwon$^{1}$,~~D.~D. Tolla$^{1,2}$\\[3mm]
{\it $^{1}$Department of Physics,~BK21 Physics Research Division,
 Autonomous Institute of Natural Science,~Institute of Basic Science, Sungkyunkwan University, Suwon 16419, Korea\\
$^{2}$University College,
Sungkyunkwan University, Suwon 16419, Korea}\\[2mm]
{\it ~yoonbai@skku.edu,~okab@skku.edu,~ddtolla@skku.edu}}

\end{center}
\vspace{15mm}

\begin{abstract}
We  construct the 4-dimensional ${\cal N}=\frac12$  and ${\cal N}=1$ inhomogeneously mass-deformed super Yang-Mills  theories from the ${\cal N} =1^*$ and   ${\cal N} =2^*$ theories, respectively, and analyse their supersymmetric vacua.  The inhomogeneity is attributed to  the dependence of background fluxes in the type IIB supergravity  on a single spatial coordinate.  This gives rise to inhomogeneous mass functions in the ${\cal N} =4$ super Yang-Mills theory which describes the dynamics of D3-branes.  
The Killing spinor equations for those inhomogeneous theories lead to the supersymmetric vacuum equation and a boundary condition. We investigate two types of solutions in the $ {\cal N}=\frac12$ theory, corresponding to the cases of asymptotically constant mass
functions and periodic mass functions. For the former case, the boundary condition gives a relation between the parameters of two possibly distinct vacua at the asymptotic boundaries. Brane interpretations for corresponding vacuum solutions in type IIB supergravity are also discussed. For the latter case,  we obtain explicit forms of the periodic vacuum solutions. 
\end{abstract}

\newpage
\tableofcontents

\section{Introduction}

The original Janus solution of the type IIB supergravity was introduced as extension of an asymptotically AdS$_5 \times S^5$ solutions, by allowing the variation of the dilaton field along one spatial direction~\cite{Bak:2003jk}. The dilaton variation can approaches two distinct constant values at the two boundaries of the chosen space direction, creating a codimension one defect in between, which is usually referred to as an interface. Though they proved to be stable, those solutions are non-supersymmetric. However, some variants of these solutions which preserve some supersymmetry were found later~\cite{ Clark:2004sb, DHoker:2007zhm, Clark:2005te, Suh:2011xc,Bobev:2019jbi,  Bobev:2020fon}. The holographic dual field theories to those Janus solutions are  the 4-dimensional ${\cal N} = 4$ super Yang-Mills (SYM) gauge theories on each side of the interface with space dependent gauge coupling, which varies across the interface~\cite{DHoker:2006qeo,Kim:2008dj, Gaiotto:2008sd, Hashimoto:2014vpa, Choi:2017kxf}. At the two boundaries of the one spatial direction, the values of the gauge coupling correspond to the asymptotic values of the dilaton field at those boundaries.

In the ${\cal N} = 4$ SYM  theory, the  gauge coupling  is the only parameter of the theory, which can have a non-trivial space dependence. On the other hand, in the dual gravity theory, in addition to or instead of the dilaton variation, one can consider the spatial variation of the background $p$-form fluxes  to explore more general extensions of the asymptotically AdS$_5 \times S^5$ solutions. In this case, the dual field theories with some  background form fluxes are the mass-deformed super Yang-Mills (mSYM) gauge theories~\cite{Vafa:1994tf, Donagi:1995cf, Polchinski:2000uf}, which are refereed as the ${\cal N} = 1^*$ and ${\cal N} = 2^*$ theories. The gauge/gravity duality of those theories were investigated, $e.g.$ see \cite{Freedman:1999gp, Girardello:1999bd, Polchinski:2000uf, Aharony:2000nt, Bobev:2018eer, Petrini:2018pjk, Bobev:2019wnf}.  When the background form fluxes are a space coordinate dependent, the mass parameters in the mSYM theories  are inhomogeneous along that spatial direction. This generalization of the mass deformation is  analogous to that of the ABJM theory~\cite{Aharony:2008ug}. Recently, the ${\cal N} = 3$ supersymmetric inhomogeneous gauge model with space-dependent mass parameters were constructed in \cite{Kim:2018qle} in the context of the ${\cal N}=6$  mass-deformed ABJM theory~\cite{Hosomichi:2008jb, Gomis:2008vc}. Less supersymmetric generalization and vacuum solutions of the ${\cal N}=3$ inhomogeneously mass-deformed ABJM (ImABJM) model were 
studied in \cite{Kim:2019kns}. See also \cite{DHoker:2009lky} for the Janus deformation of the ABJM theory. A general discussion for the Janus couplings in various field theories, including the mass parameters, was given in \cite{Anderson:2019nlc}. Holographic duality descriptions for the inhomogeneous (spatially modulated) mass deformations, which preserve 1/4 supersymmetry of the $d=11$ supergravity, were constructed in \cite{Gauntlett:2018vhk} in terms of the Q-lattice construction~\cite{Donos:2013eha}. The construction in \cite{Gauntlett:2018vhk} was also generalized in \cite{Arav:2018njv}. Based on the $Q$-lattice construction, finite temperature generalizations of the gravity dual for the ${\cal N} = 3$ ImABJM model were constructed in \cite{Ahn:2019pqy,Hyun:2019juj}, where the black brane,  AdS soliton solutions~\cite{Ahn:2019pqy},  and thermodynamic properties~\cite{ Hyun:2019juj} were discussed. It was also reported that, as an extension of \cite{Bobev:2013yra}, there are RG interface solutions~\cite{Arav:2020asu}, which have the ABJM theory on one side of the interface and one of the two $G_2$ invariant ${\cal N} =1$ $d=3$ superconformal field theories in $d=4$ supergravity. See also refs. \cite{Gutperle:2012hy, Herzog:2019bom, Kim:2020unz} for spatially dependent
relevant operators.

In this paper, we construct supersymmetric models by  inhomogeneous extensions of the ${\cal N}=1^*$ and ${\cal N}=2^*$ mSYM theories.  The homogeneous mass-deformations arise from magnetic RR 3-form fluxes, which are dual to electric RR 7-form fluxes interacting  with the D3-branes by the process of the Myers dielectric effect~\cite{Myers:1999ps}. Then, the inhomogeneously mass-deformed super Yang Mills (ImSYM)    model corresponds to the case where this background 7-form fluxes are space dependent.
We show that when the mass parameters  in mSYM theories are inhomogeneous along one spatial direction, half of the supersymmeries are preserved if appropriate additional scalar fields mass terms are added  to the Lagrangian of the homogeneous mSYM 
theories.\footnote{Similar constructions preserving half supersymmetries~\cite{Kim:2018qle, Kim:2019kns} were obtained from the ${\cal N} = 2,3,6 $ mABJM 
theories~\cite{Hosomichi:2008jb, Gomis:2008vc,Kim:2012gz}.} 
As a result, we obtain the  ${\cal N}=\frac12$  and ${\cal N}=1$ ImSYM    models from the ${\cal N}=1^*$ and ${\cal N}=2^*$ mSYM theories, respectively. These inhomogeneous constructions were also given in 
\cite{Arav:2020obl} using a different method. Dual gravity solutions and gauge/gravity duality correspondence in terms of the holographic renormalization method were also constructed  \cite{Arav:2020obl} in $d=5$ supergravities ,   which arise from consistent truncations of type IIB supergravity. In \cite{Arav:2020asu}, the RG interface solutions joining the ${\cal N} = 4$ SYM theory and the ${\cal N} =1$ Leigh-Strassler superconformal field theory~\cite{Leigh:1995ep} was constructed, as a dual gravity solution for the ${\cal N} = \frac{1}{2}$ ImSYM model with one non-vanishing mass function.

In the $ {\cal N}=\frac12$ ImSYM model with  all the masses are non-vanishing, we investigate two types of vacuum solutions corresponding to the cases of asymptotically constant mass functions and periodic mass functions.    In the 4-dimensional ${\cal N} = 1$  language, the ${\cal N} = 4$ SU($N$) SYM theory consists of one vector multiplet and three chiral multiplets, all in the $N$-dimensional adjoint representations of the gauge group.  In the absence of the inhomogeneity, the ${\cal N}=1^*$  mSYM theory is obtained by adding a mass deformation with the three chiral multiplets assigned different mass parameters, whereas for the ${\cal N}=2^*$  case one chiral multiplet remains massless and the other two assigned equal mass parameters. For  the ${\cal N}=1^*$ theory with all the three mass parameters are non-zero, the supersymmetric vacuum solutions are the Higgs branch solutions, which are expressed in terms of $N$-dimensional, generally reducible, representations of the $su(2)$ algebra~\cite{Polchinski:2000uf}. However, for  the ${\cal N}=1^*$ theory with some of the mass parameters are zero and for the    ${\cal N}=2^*$ theory, the vacuum solutions are in the Coulomb branch. Here, we focus on the  ${\cal N}=\frac12$  ImSYM model, which corresponds to the former case.   From the Killing spinor equation of the  ${\cal N}=\frac12$  ImSYM model, we obtain the supersymmetric vacuum equations and the boundary condition that the inhomogeneous solutions should satisfy at the asymptotic boundaries of the one spacial coordinate.  We discuss two types of  solutions, corresponding to the cases of asymptotically constant mass functions  and periodic mass functions. In the former case, we find the relations which are necessary for the solutions to satisfy the asymptotic boundary condition. We discuss brane configurations corresponding to the vacuum solutions. For the later case, the boundary condition is trivially satisfied and we obtain explicit forms of the periodic solutions.  

The remaining parts of this paper are summarized as follows. In section 2, we review some necessary aspects of the ${\cal N}=1^*$ and ${\cal N}=2^*$ mSYM    theories and then construct the corresponding ${\cal N}=\frac12$  and ${\cal N}=1$ ImSYM    models by allowing the variation of the mass parameters along one spatial direction. We present the detailed procedure for the ${\cal N}=\frac12$ case and briefly summarize the case of ${\cal N}=1$ in the Appendix. In section 3, we obtain  the supersymmetric vacuum equation and the  boundary conditions for the ${\cal N}=\frac12$ ImSYM model from its Killing spinor equations. We investigate the boundary condition for the vacuum solutions with asymptotically  constant mass functions, discuss corresponding brane configurations, and explicitly construct the vacuum solutions associated with periodic mass functions.  In section 4, we draw our conclusions. 
\\

\noindent
{\bf Note Added:} While this paper was being completed, a paper \cite{Arav:2020obl}
appeared, which also constructs the supersymmetric inhomogeneously mass-deformed SYM models.

\section{${\cal N} = 1$ and ${\cal N} = \frac{1}{2}$ ImSYM Models}
The mass-deformation of the 4-dimensional ${\cal N}=4$ SYM theory breaks the supersymmetry, partially or totally, depending on the choice of the bosonic and fermionic mass matrices.
In this section, we construct supersymmetric theories by allowing  the  mass parameters  in the ${\cal N}=1^*$ and ${\cal N}=2^*$ mSYM theories to be inhomogeneous. We  show that when the inhomogeneity of the mass parameters is only along one spatial direction, the resulting ImSYM models with appropriate interaction terms preserve half of the ${\cal N}=1^*$ and ${\cal N}=2^*$ supersymmetries.

\subsection{Review of ${\cal N}=1^*$ and ${\cal N}  =2^* $ mSYM theories}
In this subsection, we review the SU($N$) mSYM theories~\cite{Vafa:1994tf,Donagi:1995cf}, known as the ${\cal N}=1^*$ and ${\cal N}=2^*$ theories. Dual supergravity theories of those theories were studied in \cite{Girardello:1999bd,Polchinski:2000uf}.  Though the ${\cal N}=1$ superfield formalism is widely used to discuss the mSYM theories with constant mass parameters, in order to introduce the inhomogeneous mass deformations, we found that the component field approach is more convenient. Therefore, we begin by writing the ${\cal N}=4$  SYM theory action, which can be obtained from the dimensional reduction of the 10-dimensional ${\cal N}=1$ SYM theory, by using the component fields in the adjoint representations of SU($N$) as 
\begin{align}\label{4dact}
{\cal  L_{\rm SYM}}=&{\rm tr}\left[-\frac12F_{\alpha\beta}F^{\alpha\beta}-D^\alpha\phi_a
D_\alpha\phi_a+\frac{g^2}2\,[\phi_a,\phi_b]^2+ i\bar\psi_p\gamma^\alpha 
D_\alpha\psi_p-{g}\Big(\bar\psi_p\big(\Gamma_a^{pq}P_{+}+\bar\Gamma_a^{pq}P_{-}
\big)[\phi_a,\psi_q]\Big)\right],
\end{align}
where $P_{\pm}=\frac{1\pm\gamma_5}2$ are the 4-dimensional chirality projection operators, $\alpha,\beta=0,...,3$ are the 4-dimensional space-time indices, $a,b,c=1,...,6$ and, $p,q,r=1,...,4$, respectively, are the indices of the $SO(6)$ and $SU(4)$ representations of the global symmetries. The  $\psi_p$'s with $\bar\psi_p=\psi_p^T\gamma^0$
 are Majorana fermions and $\phi_a$'s are Hermitian scalar fields.  We also introduced $\Gamma_a^{pq}=v^+_p\Gamma_av^+_q$ and
  $\bar\Gamma_a^{pq}=v^-_p\Gamma_av^-_q$, where $\Gamma_a$ are the six-dimensional Euclidean space gamma matrices, whereas
$v^+_p$ and $v^-_p$ are the eigenvectors of
  $\Gamma_\ast=-i\Gamma_1...\Gamma_6$ with eigenvalues +1, -1,
  respectively.    The Clifford algebra for the 4-dimensional gamma matrices is given by:
 $\{\gamma^\alpha,\gamma^\beta\}=2\eta^{\alpha\beta}$ with the signature $(-,+,+,+)$.
 
The  ${\cal N}=4$ supersymmetry transformation rules are also obtained from the corresponding 10-dimensional ${\cal N}=1$ supersymmetry transformations, and they are
given by \begin{align}\label{N=4susy}
 &\delta_\epsilon
 A_{\alpha}=\tcr{}i\bar\psi_p\gamma_\alpha\epsilon_p,
 \qquad\delta_\epsilon
 \phi_a=-i\bar\psi_p\big(\Gamma_a^{pq}P_{+}+\bar\Gamma_a^{pq}P_{-}
\big)\epsilon_q,\nn\\
&\delta_\epsilon\psi_p=iF_{\alpha\beta}\Sigma^{\alpha\beta}\epsilon_p+
 \gamma^{\alpha}D_\alpha\phi_a\big(\Gamma_a^{pq}P_{+}
+\bar\Gamma_a^{pq}P_{-}\big)\epsilon_q-{g}[\phi_a,
\phi_b]\big(\Gamma^{pq}_{ab}P_{+}
+\bar\Gamma^{pq}_{ab}P_{-}\big)\epsilon_q,
 \end{align}
where $\Sigma^{\alpha\beta}=\frac i4[\gamma^\alpha,\gamma^\beta]$, the supersymmetry parameters $\epsilon_p$'s are Majorana fermions, and $\Gamma_{ab}^{pq}=v^-_p\Sigma^{ab}v^+_q$,~
$\bar\Gamma^{ab}_{pq}=v^+_p\Sigma^{ab}v^-_q$ with $\Sigma^{ab} = -\frac{i}{4}[\Gamma^a,\,\Gamma^b]$.
By rearranging  those, one can write
 \begin{align}
 &\Gamma^{pq}_{ab}=\frac{i}{4}\big(\bar\Gamma_a^{pr}\Gamma_b^{rq}
 -\bar\Gamma_b^{pr}\Gamma_a^{rq}\big),\qquad\qquad \bar\Gamma^{pq}_{ab}=\frac{i}{4}\big(\Gamma_a^{pr}\bar\Gamma_b^{rq}
 -\Gamma_b^{pr}\bar\Gamma_a^{rq}\big).
 \end{align}
We also obtain the explicit forms of  the matrix elements of $\Gamma_a$ from some choice of the 10-dimensional gamma matrices,
\begin{align}\label{Gamma}
&\Gamma_1^{pq}=i\big(\delta_{p1}\delta_{q4}
-\delta_{p4}\delta_{q1}+\delta_{p2}\delta_{q3}
-\delta_{p3}\delta_{q2}\big),\qquad \Gamma_2^{pq}=i\big(\delta_{p1}\delta_{q2}
-\delta_{p2}\delta_{q1}+\delta_{p3}\delta_{q4}
-\delta_{p4}\delta_{q3}\big),\nonumber\\
&\Gamma_3^{pq}=i\big(\delta_{p1}\delta_{q3}
-\delta_{p3}\delta_{q1}-\delta_{p2}\delta_{q4}
+\delta_{p4}\delta_{q2}\big),\qquad\Gamma_4^{pq}=-\big(\delta_{p1}\delta_{q4}
-\delta_{p4}\delta_{q1}-\delta_{p2}\delta_{q3}
+\delta_{p3}\delta_{q2}\big),\nonumber\\
&\Gamma_5^{pq}=\big(\delta_{p1}\delta_{q2}
-\delta_{p2}\delta_{q1}-\delta_{p3}\delta_{q4}
+\delta_{p4}\delta_{q3}\big),\qquad\Gamma_6^{pq}=-\big(\delta_{p1}\delta_{q3}
-\delta_{p3}\delta_{q1}+\delta_{p2}\delta_{q4}
-\delta_{p4}\delta_{q2}\big),
\end{align}
and  $\bar\Gamma_a^{pq}=-\Gamma_a^{pq},~{\rm for}~ a=1,2,3$ and $\bar\Gamma_a^{pq}=\Gamma_a^{pq},~{\rm for}~ a=4,5,6$. These matrix representations of the gamma matrices satisfy the Clifford algebra: $\bar\Gamma_a^{pr}\Gamma_b^{rq}
 +\bar\Gamma_b^{pr}\Gamma_a^{rq}=\Gamma_a^{pr}\bar\Gamma_b^{rq}
 +\Gamma_b^{pr}\bar\Gamma_a^{rq}=-2\delta_{ab}\delta^{pq}$. With the help of this algebra and the Fierz identities, one can show that the Lagrangian in \eqref{4dact} is invariant under the ${\cal N}=4$ supersymmetry transformations in \eqref{N=4susy}. 

Depending on the choice of the mass matrices, the mass-deformation of \eqref{4dact}  preserves only one or two of the ${\cal N}=4$ supersymmetries. The resulting theories are the ${\cal N}=1^*$ and ${\cal N}=2^*$ mSYM theories. Here we mainly review the ${\cal N}=1^*$ theory and summarize the ${\cal N}=2^*$ case in the Appendix. Without loss of generality, in the ${\cal N}=1^*$  mSYM theory, we  choose $\epsilon_4=\epsilon$ as the parameter of the 
unbroken supersymmetry with the other three parameters $(\epsilon_1,\epsilon_2,\epsilon_3)$ set to zero. This means we can write
\begin{align}\label{ep4}
\epsilon_p = \delta_{p4} \epsilon. 
\end{align} Then the supersymmetry
transformation rules in \eqref{N=4susy} are reduced to
\begin{align}\label{N=2*susy}
 &\delta_\epsilon
 A_{\alpha}=\tcr{}i\bar\psi_4\gamma_\alpha\epsilon,\qquad \delta_\epsilon
 \phi_a=-i\bar\psi_p\big(\Gamma_a^{p4}P_{+}+\bar\Gamma_a^{p4}P_{-}
\big)\epsilon,\\
 &\delta_\epsilon\psi_p=
 iF_{\alpha\beta}\Sigma^{\alpha\beta}\delta_{p4}\epsilon+\gamma^{\alpha}D_\alpha\phi_a\big(\Gamma_a^{p4}P_{+}
 +\bar\Gamma_a^{p4}P_{-}\big)\epsilon-{g}[\phi_a,
 \phi_b]\big(\Gamma^{p4}_{ab}P_{+}
 +\bar\Gamma^{p4}_{ab}P_{-}\big)\epsilon.\nonumber
  \end{align}
  In the presence of the mass-deformation, the total action will be invariant if the supersymmetry variations of the fermionic fields are modified by including the following additional transformations, \begin{align}\label{delta'}
\delta'_\epsilon\psi_p=\mu_{pq}\phi_a\big(\Gamma_a^{q4}P_{+}
 +\bar\Gamma_a^{q4}P_{-}\big)\epsilon,
\end{align}
for specific fermionic mass matrix $\mu_{pq}$, whereas $\delta_\epsilon'A_\alpha=\delta_\epsilon'\phi_a=0$.  In this subsection, we assume  $\mu_{pq}$ is constant.  With the modification in \eqref{delta'}, the variation of the undeformed Lagrangian in \eqref{4dact} is rearranged as 
\begin{align}\label{deltaL}
(\delta_\epsilon+\delta'_\epsilon){\cal  L_{\rm SYM}}={\rm tr}\big(2i\mu_{pq}\bar\psi_p(\delta_\epsilon+\delta'_\epsilon)\psi_q+2M_{ab}\phi_a\delta_\epsilon\phi_b
-3igT_{abc}[\phi_b,\,\phi_c]\delta_\epsilon\phi_a\big),
\end{align}
where
 $\mu_{pq}={\rm diag}(\mu_1,\mu_2,\mu_3,0)$, the bosonic mass matrix $ M_{ab}={\rm
diag}(\mu_1^2,\mu_3^2,\mu_2^2,\mu_1^2,\mu_3^2,\mu_2^2)$
with constant mass parameters $\mu_{m=1,2,3}$, and the nonvanishing components of the antisymmetric three-form tensor $T_{abc}$ are
\begin{align}\label{threeTen}
&T_{234}=\frac13(\mu_1-\mu_2-\mu_3),\quad
T_{126}=\frac13(\mu_1-\mu_2+\mu_3),\nonumber\\
&T_{135}=\frac13(\mu_1+\mu_2-\mu_3),\quad T_{456}=\frac13(\mu_1+\mu_2+\mu_3).
\end{align} As we mentioned above, in this subsection, we assume $\mu_m$ are constant.
Therefore, if we add the following mass-deformation to the Lagrangian
in \eqref{4dact}, 
\begin{align}\label{4dmdact}
{\cal L}_\mu={\rm tr}\big(-i\mu_{pq} \bar\psi_p\psi_q-M_{ab}\phi_a\phi_b
+igT_{abc}\phi_a[\phi_b,\,\phi_c]\big),
\end{align}
then the total Lagrangian is invariant
of the  ${\cal N}=1$ supersymmetry transformation, i.e.
\begin{align}\label{N=2susy}
(\delta_\epsilon+\delta'_\epsilon)({\cal  L_{\rm SYM}}+{\cal L}_\mu)=0.
\end{align}
Here, we would like to note that the ${\cal N}=2^*$ homogeneous mSYM theory can be obtained from the ${\cal N}=1^*$ theory by setting $\mu_1 = \mu_2=\mu$,  $\mu_3 = 0$ and choosing the unbroken supersymmetry parameters as $\epsilon_p=\delta_{pi}\epsilon_i$ with $i=\{3,4\}$. See the Appendix for the details.

\subsection{Construction of the ${\cal N} = \frac12$ and ${\cal N} = 1$ ImSYM models }
In this subsection, we  consider the inhomogeneous extension of the mSYM theory summarized in the previous subsection by assuming that the mass parameters are coordinate dependent$\big(\mu_m=\mu_m(x^\alpha)\big)$. In that case, the supersymmetry variation of the total Lagrangian in \eqref{N=2susy}, which we can expand as
\begin{align}\label{N=2susy-a}
(\delta_\epsilon+\delta'_\epsilon)({\cal  L_{\rm SYM}}+{\cal L}_\mu)=\delta'_\epsilon{\cal  L_{\rm SYM}}+\delta_\epsilon {\cal L}_\mu+\delta'_\epsilon{\cal L}_\mu,
\end{align}
is not zero, where we have used the fact $\delta_\epsilon{\cal  L_{\rm SYM}}=0$. The first term on the right hand side of \eqref{N=2susy-a} is expressed as \begin{align}\label{deltapL}
\delta'_\epsilon{\cal  L_{\rm SYM}} &={\rm tr}\big[2i(\partial_\alpha\mu_m)\bar\psi_m\gamma^\alpha \phi_a\big(\Gamma_a^{m4}P_{+}
 +\bar\Gamma_a^{m4}P_{-}\big)\epsilon+2i\mu_m\bar\psi_m\gamma^\alpha D_\alpha\phi_a\big(\Gamma_a^{m4}P_{+}
 +\bar\Gamma_a^{m4}P_{-}\big)\epsilon \nn\\
 &~~~-2g\mu_m\bar\psi_p[\phi_a,\phi_b]\big(\Gamma_a^{pm}\Gamma_b^{m4}P_{+}+\bar\Gamma_a^{pm}\bar\Gamma_b^{m4}P_{-}
\big)\epsilon\big],
\end{align}
where $m=\{1,2,3\}$ and $p=\{1,2,3,4\}$. Similarly, the second and the third terms on the right hand side of \eqref{N=2susy-a} are given by
\begin{align}
\delta_\epsilon{\cal L}_\mu&={\rm tr}\big[-2i\mu_m D_\alpha\phi_a\bar\psi_m\gamma^{\alpha}\big(\Gamma_a^{m4}P_{+}
 +\bar\Gamma_a^{m4}P_{-}\big)\epsilon-g\mu_m [\phi_a,
 \phi_b]\bar\psi_m\big(\bar\Gamma^{mp}_{a} \Gamma^{p4}_bP_{+}
 +\Gamma^{mp}_{a}\bar\Gamma^{p4}_bP_{-}\big) \epsilon 
\nn\\
 &~~~+2iM_{ab}\phi_a\bar\psi_p(\Gamma_b^{p4}P_{+} +\bar\Gamma_b^{p4}P_{-}
\big)\epsilon+3gT_{abc}[\phi_b,\phi_c]) \bar\psi_p(\Gamma_a^{p4}P_{+}+\bar\Gamma_a^{p4}P_{-}
\big)\epsilon \big],
\nn \\
\label{deltapLm} \delta'_\epsilon{\cal L}_\mu&={\rm tr}\big[-2i\mu_{pr}\mu_{rq}\phi_a\bar\psi_p\big(\Gamma_a^{q4}P_{+}
 +\bar\Gamma_a^{q4}P_{-}\big)\epsilon\big].
\end{align}
Combing the expressions in \eqref{deltapL} and \eqref{deltapLm}, we obtain
\begin{align}\label{N=2susy-c}
(\delta_\epsilon+\delta'_\epsilon)({\cal  L_{\rm SYM}}+{\cal L}_\mu)&=2i(\partial_\alpha\mu_m){\rm tr}\Big[\Big(-\sum_{a=1}^{3}+\sum_{a=4}^{6} \Big) \phi_a\bar\psi_m \big(\Gamma_a^{m4}P_{+}
 +\bar\Gamma_a^{m4}P_{-}\big)\Big] \gamma^\alpha\epsilon.
\end{align}
This means that the mass-deformed theory can not preserve the full ${\cal N}=1$ supersymmetry when the mass parameters are inhomogeneous. 

Now we determine the coordinate dependence of the mass parameter for which half of the ${\cal N}=1$ supersymmetry is preserved. Since from \eqref{Gamma} we see that $\Gamma_a^{44}=0$, we can rewrite \eqref{N=2susy-c} as
\begin{align}
(\delta_\epsilon+\delta'_\epsilon)\big({\cal  L_{\rm SYM}}+{\cal L}_\mu\big)&=2i(\partial_\alpha\mu_p) {\rm tr}\Big[\Big(-\sum_{a=1}^{3}+\sum_{a=4}^{6}\Big)\phi_a\bar\psi_p \big(\Gamma_a^{p4}P_{+}
 +\bar\Gamma_a^{p4}P_{-}\big) \Big]\gamma^\alpha\epsilon \nn\\
 &= {\rm tr}\big[-2i( \partial_\alpha J_{ab})\phi_a\bar\psi_p \big(\Gamma_b^{p4}P_{+}+\bar\Gamma_b^{p4}P_{-}\big)\gamma^\alpha\epsilon \big],
 \end{align}
 where we have introduced 
 \begin{align}\label{Jab}
 J_{ab}={\rm diag}(\mu_1,\mu_3,\mu_2,-\mu_1,-\mu_3,-\mu_2).
\end{align}
 In order to proceed, we assume the mass parameters are inhomogeneous only along one spatial direction $x\equiv x^1$, then we obtain 
\begin{align}
&(\delta_\epsilon+\delta'_\epsilon)\big({\cal  L_{\rm SYM}}+{\cal L}_\mu\big)={\rm tr} \big[-2i J'_{ab}\phi_a\bar\psi_p \big(\Gamma_b^{p4} P_{+}+\bar\Gamma_b^{p4}P_{-} \big) \gamma^1 \epsilon \big],
\end{align}
where $J'_{ab}=\frac{d}{dx}J_{ab}$. Imposing the projection,
\begin{align}\label{pjcon}
\gamma^1\epsilon=\epsilon,
\end{align} 
which breaks half of the ${\cal N}=1$ supersymmetry, we obtain 
\begin{align}
&(\delta_\epsilon+\delta'_\epsilon)\big({\cal  L_{\rm SYM}}+{\cal L}_\mu\big)={\rm tr}\Big[2J'_{ab} \phi_a\Big(-i\bar\psi_p \big(\Gamma_b^{p4} P_{+}+\bar\Gamma_b^{p4}P_{-}\big)\epsilon \Big) \Big] ={\rm tr}\big[2 J'_{ab}\phi_a \delta\phi_b \big].
\end{align}
Therefore, we can cancel this term and obtain ${\cal N}=\frac12$ ImSYM  model by introducing an additional  mass term for the scalar field as
\begin{align}
{\cal  L}_J=-{\rm tr} \big( J'_{ab}\phi_a\phi_b \big).
\end{align} 
Similarly, the ${\cal N}=1$ ImSYM model can be obtained from the ${\cal N}=2^*$ homogeneous mSYM theory by considering a inhomogeneous mass parameter $\mu=\mu(x)$ and adding the inhomogeneous mass term ${\cal L}_J=-{\rm tr}\big(J'_{ab}\phi_a\phi_b\big)$, where $J_{ab}={\rm diag}(\mu,0,\mu,-\mu,0,-\mu)$.
See the Appendix.

\section{Vacuum Solutions of the ${\cal N} = \frac12$ ImSYM Model}

In this section, we focus on the vacuum solutions for the  ${\cal N} = \frac12$ ImSYM model, leaving the more challenging case of   the ${\cal N}=1$ ImSYM model, which includes the vacuum moduli of one massless complex scalar field.

\subsection{Vacuum equations and the boundary condition}

  In order to find  vacuum solutions for the  ${\cal N} = \frac12$ ImSYM model, we set the fermionic fields and the gauge fields to zero,  then the energy-momentum tensor from the action $S = \int d^4x \left( {\cal  L_{\rm SYM}} + {\cal L}_\mu + {\cal L} _J \right) $ is given by 
\begin{align} \nn
T_{\mu\nu}={\rm tr}\Big[2\partial_{\mu} \phi_a \partial_{\nu} \phi_a+g_{\mu\nu} \Big(-\partial^{\alpha} \phi_a
\partial_{\alpha}\phi_a+\frac{g^2}2\,[\phi_a,\phi_b]^2-(M_{ab}+ J'_{ab})\phi_a\phi_b
+igT_{abc}\phi_a[\phi_b,\,\phi_c] \Big)\Big].
\end{align}
For the inhomogeneous mass deformation along the single spatial direction $x$, the vacuum solutions are specified by the inhomogeneous scalar fields $\phi_a=\phi_a(x)$. Then the vacuum energy is given by
\begin{align}\label{vac-E}
E_0=\int d^3x~T_{00}=\int d^3x~{\rm tr} \Big[\phi_a'
\phi_a'-\frac{g^2}2\,[\phi_a,\phi_b]^2+(M_{ab}+ J'_{ab})\phi_a\phi_b
-igT_{abc}\phi_a[\phi_b,\,\phi_c]\Big].
\end{align}

Next we show that the vacuum energy in \eqref{vac-E} can be rewritten as the square of the norm of the fermionic variations $|(\delta_\epsilon+\delta'_\epsilon)\psi_p|^2$ plus some total derivative with respect to $x$. Similar construction was also given in the ${\cal N} = 3$ ImABJM 
models~\cite{Kim:2019kns}.  Using \eqref{ep4},  we  write the total fermionic variation as
\begin{align}
  (\delta_\epsilon+\delta'_\epsilon)\psi_p=
 &iF_{\alpha\beta}\Sigma^{\alpha\beta}\delta_{p4}\epsilon+D_\alpha\phi_a\big(\Gamma_a^{p4}P_-
 +\bar\Gamma_a^{p4}P_+\big)\gamma^{\alpha}\epsilon-{g}[\phi_a,
 \phi_b]\big(\Gamma^{p4}_{ab}P_+
 +\bar\Gamma^{p4}_{ab}P_-\big)\epsilon\nonumber\\
 &+\mu_{pq}\phi_a\big(\Gamma_a^{q4}P_+
 +\bar\Gamma_a^{q4}P_-\big)\epsilon.
  \end{align}
  Since $A_\alpha=0$ and  $\phi_a=\phi_a(x)$ at the vacuum, this is rewritten as  
 \begin{align}\label{delta-psi}
  (\delta_\epsilon+\delta'_\epsilon)\psi_p=
 &\Big[\phi_a'\big(\Gamma_a^{p4}P_-
 +\bar\Gamma_a^{p4}P_+\big)-{g}[\phi_a,
 \phi_b]\big(\Gamma^{p4}_{ab}P_+
 +\bar\Gamma^{p4}_{ab}P_-\big)+\mu_{pq}\phi_a \big(\Gamma_a^{q4}P_+
 +\bar\Gamma_a^{q4}P_-\big)\Big]\epsilon,
\end{align}
where we have also used the projection  condition \eqref{pjcon} to obtain the first term on the right hand side.

  After some algebra, the trace of the norm for the square bracket in \eqref{delta-psi} can be written as 
\begin{align}\label{deltapsi2}
 &{\rm tr}\Big [\left|\phi_a'\big(\Gamma_a^{p4}P_-
 +\bar\Gamma_a^{p4}P_+\big)-{g}[\phi_a,
 \phi_b]\big(\Gamma^{p4}_{ab}P_+
 +\bar\Gamma^{p4}_{ab}P_-\big)+\mu_{pq}\phi_a\big(\Gamma_a^{q4}P_+
 +\bar\Gamma_a^{q4}P_-\big)\right|^2 \Big]\\
 &={\rm tr}\Big[\phi_a'\phi_a'-\frac{ig}3\Big(\tilde T_{abc}\phi_a[\phi_b,\phi_c]\Big)'- J_{ab}\Big(\phi_a\phi_b\Big)'-\frac{g^2}2[\phi_a,\phi_b]^2-igT_{abc}\phi_a[\phi_b, \phi_c]+M_{ab}\phi_a\phi_b\Big],\nn
 \end{align}
where $|A|^2\equiv AA^\dagger$, we have used  $\Gamma^{pq\ast}_a=\bar\Gamma^{pq}_a$, and  introduced the constant antisymmetric tensor $\tilde T_{abc}$ with the only non zero elements,
\begin{align}\label{tilT}
\tilde T_{126}=\tilde T_{135}=-\tilde T_{234}=-\tilde T_{456}=-1.
\end{align}  
Combining \eqref{deltapsi2} with the vacuum energy in \eqref{vac-E} gives the expected result:  
\begin{align}\label{E0}
 E_0&=\int d^3x~{\rm tr}\Big[\left|\phi_a'\big(\Gamma_a^{p4}P_-
 +\bar\Gamma_a^{p4}P_+\big)-{g}[\phi_a,
 \phi_b]\big(\Gamma^{p4}_{ab}P_+
 +\bar\Gamma^{p4}_{ab}P_-\big)+\mu_{pq}\phi_a\big(\Gamma_a^{q4}P_+
 +\bar\Gamma_a^{q4}P_-\big)\right|^2\Big] \nn\\
&~~~+\int d^3 x~{\cal K}',
 \end{align}
 where the total derivative term is 
 \begin{align}\label{calK}
{\cal K} = {\rm tr}\Big(J_{ab}\phi_a\phi_b+\frac{ig}{3}\tilde T_{abc}\phi_a[\phi_b,\phi_c]\Big). 
\end{align}
Then the vacuum equation becomes  
\begin{align}\label{vacEq-1}
 &\phi_a'\big(\Gamma_a^{p4}P_-
 +\bar\Gamma_a^{p4}P_+\big)-{g}[\phi_a,
 \phi_b]\big(\Gamma^{p4}_{ab}P_+
 +\bar\Gamma^{p4}_{ab}P_-\big) +\mu_{pq}\phi_a \big(\Gamma_a^{q4}P_+
 +\bar\Gamma_a^{q4}P_-\big)=0. 
 \end{align}
 Projecting \eqref{vacEq-1} by $P_\pm$,  we can split it  into two simpler equations, 
 \begin{align}\label{pjsim}
 &\phi_a'\bar\Gamma_a^{p4}-{g}[\phi_a,
 \phi_b]\Gamma^{p4}_{ab}+\mu_{pq}\phi_a\Gamma_a^{q4}=0,\qquad\phi_a'\Gamma_a^{p4}-{g}[\phi_a,
 \phi_b]\bar\Gamma^{p4}_{ab}+\mu_{pq}\phi_a\bar\Gamma_a^{q4}=0.
  \end{align}
 The second equation in \eqref{pjsim}  is the complex conjugate of the first one.  Therefore, the supersymmetric condition $E_0=0$ in \eqref{E0} leads to the two relations,   
\begin{align}\label{vacEq5}
 &\phi_a'\bar\Gamma_a^{p4}-\frac{ig}2[\phi_a,
 \phi_b]\bar\Gamma^{pr}_{a}\Gamma^{r4}_{b} +\mu_{pq}\phi_a\Gamma_a^{q4}=0, 
\\\label{bdc2}
& \int_{x_L}^{x_R} dx\,\, {\cal K}' = 0 \quad \Longleftrightarrow \quad {\cal K} |_{x\to x_L} = {\cal K} |_{x\to x_R},
\end{align}
where $x_L$ and $x_R$ represent the asymptotic boundaries along the $x$-direction. 
 More explicitly, by inserting $p=1,\cdots,4 $ in \eqref{vacEq5}, we obtain the following 4  equations:
 \begin{align}\label{vaceq6}
& i\phi_1'+\phi_4'- g\Big( i\big([\phi_2,\phi_3]+[\phi_5,\phi_6]\big)+\big( [\phi_2,\phi_6]+[\phi_3,\phi_5]\big) \Big)- \mu_1\big(i\phi_1-\phi_4\big)=0, \nn \\
&i\phi_3'-\phi_6'+g\Big(-i\big([\phi_1,\phi_2]-[\phi_4,\phi_5]\big)+\big( [\phi_1,\phi_5]-[\phi_2,\phi_4]\big)\Big)-\mu_2\big(i\phi_3 +\phi_6\big)=0,\nn \\
& i\phi_2'+\phi_5'+g\Big(i\big([\phi_1,\phi_3]+[\phi_4,\phi_6]\big)+\big([\phi_1,\phi_6]+[\phi_3,\phi_4] \big)\Big) -\mu_3\big(i\phi_2-\phi_5\big)=0, \nn \\
 &[\phi_1,\phi_4]+[\phi_2,\phi_5]-[\phi_3,\phi_6]=0.
 \end{align}
Introducing complex scalars as
 \begin{align}\label{comsc}
 \Phi_1=g(\phi_1+i\phi_4), \qquad \Phi_3 =g(\phi_2+i\phi_5),\qquad \Phi_2= g(\phi_3 -i\phi_6),
 \end{align}
 the vacuum equations in \eqref{vaceq6} are written as 
 \begin{align}\label{vacEq-2}
 &\Phi_i^{\dagger '}+\frac{1}{2} \sum_{j,k=1}^3 \epsilon_{ijk}[\Phi_j,\Phi_k]-\mu_i\Phi_i=0,
 \qquad \sum_{i=1}^3[\Phi_i,\, \Phi_i^\dagger] = 0.
 \end{align}
 
 For arbitrary mass functions $\mu_i(x)$'s $(i= 1,2,3)$,   solving the vacuum equations in \eqref{vacEq-2} is nontrivial. In this paper, we consider two kinds of mass functions,   which are  asymptotically constant mass functions,  and  periodic mass functions. For the former case, the inhomogeneous vacuum solutions may approach two different supersymmetric vacua of the ${\cal N} = 1^*$ theory at the two asymptotic boundaries of the space direction $x$. In the subsection \ref{convac}, we find a relation which   matches the  boundary condition in \eqref{bdc2} for the case  of asymptotically constant mass functions. In the subsection \ref{pvs}, we consider periodic mass functions, which trivially satisfy the boundary condition, and obtain explicit forms of  vacuum solutions.    
 
 \subsection{Boundary condition for asymptotically constant mass functions }\label{convac}
For mass functions, which are asymptotically constants, i.e., we consider the mass functions satisfying the conditions, 
\begin{align}
\lim_{x_L\to -\infty} \mu_i(x_L) = \mu_{Li}, \qquad \lim_{x_R\to \infty} \mu_i(x_R) = \mu_{Ri},\qquad (i=1,2,3)
\end{align}
with constants  $ \mu_{Li}$ and $ \mu_{Ri}$. This means that in the asymptotic region the vacuum  equations in \eqref{vacEq-2} become those of the  ${\cal N} = 1^*$ theory. Therefore, to obtain the quantity ${\cal K} |_{x\to \pm \infty}$, lets consider the vacua of the ${\cal N} = 1^*$ theory. For constant vacuum configurations,  \eqref{vacEq-2} becomes 
\begin{align}\label{cmve}
[\Phi_i,\Phi_j]-\epsilon_{ijk}(\mu_{0k}\Phi_k)=0, \qquad \sum_{i=1}^{3} [\Phi_i^\dagger,\, \Phi_i] = 0,
\end{align}
where $\mu_{0i}$'s $(i=1,2,3)$ denote constant mass parameters.
The general solution for the homogeneous vacuum configurations  is given 
by~\cite{Polchinski:2000uf}
\begin{align}\label{Phi123}
\Phi_1 = -i\sqrt{\mu_{02}\mu_{03}} \,\,T_1,\qquad \Phi_2 = -i\sqrt{\mu_{01}\mu_{03}}\,\,T_2,\qquad \Phi_3 = -i\sqrt{\mu_{01}\mu_{02}}\,\,T_3,
\end{align}
where  $T_i$'s form an $N$-dimensional (in general reducible) representation of the $su$(2) Lie algebra $[T_i,T_j] = i \epsilon_{ijk} T_k$.  
Here we follow the convention in \cite{Naculich:2001us}. Then the Hermitian generators $T_i$'s are written in the following block-diagonal form 
\begin{align}
 \label{Ti}
 T_i = \left( \begin{array}{ccc} T_i^{(n_1)}  & & \\ 
                        & \ddots &   \\
                        & & T_i^{(n_l)}
 \end{array} \right )   
\end{align}
with the constraint 
\begin{align}\label{constr1}
\sum_{k=1}^{l} n_k  =N,
\end{align}
where $T_i^{(n_k)}$'s are  the $n_k$-dimensional irreducible representation. The  $\{n_1, n_2, \cdots n_l\}$ are the partition of $N$. If $N_n$ denotes the number of $T_i^{(n)}$'s in a given representation, then one can rewrite the constraint \eqref{constr1} as
\begin{align}
\sum_{n=1}^\infty n N_n = N. 
\end{align}
The $N_n$'s are usually refereed to as the occupation numbers. The supersymmetric vacua are classified by the set of the occupation number $\{ N_n \}$.  

Comparing \eqref{comsc} and \eqref{Phi123}, we obtain the vacuum configuration for the real scalar fields, 
\begin{align}\label{phi456}
\phi_1 =\phi_2 = \phi_3 = 0, \quad \phi_4 = -\frac{\sqrt{\mu_{02}\mu_{03}}}{g} \, T_1, \quad \phi_5 = -\frac{\sqrt{\mu_{01}\mu_{02}}}{g}T_3, \quad \phi_6 =\frac{\sqrt{\mu_{01}\mu_{03}}}{g}\,T_2.
\end{align} 
Plugging \eqref{phi456}, \eqref{Jab}, and \eqref{tilT} into \eqref{calK} and using the relation \eqref{Ti}, we have 
\begin{align}\label{Kvac}
{\cal K}|_{{\rm vac}} &= -\frac{1}{3g^2} \mu_{01} \mu_{02} \mu_{03} \,{\rm tr}\left( T_1^2 + T_2^2 + T_3^2\right)
\nonumber\\
&=-\frac{1}{12g^2} \mu_{01} \mu_{02} \mu_{03} \,\sum_{n=1}^\infty n(n^2 -1) N_n , 
\end{align}
where we used the relation $(T_1^{(n)})^2+ (T_1^{(n)})^2 + (T_1^{(n)})^2 = c_2(n) \1_{n\times n}$ with the quadratic Casimir $c_2(n) = \frac{1}{4} (n^2 -1)$. 
Therefore, the boundary condition \eqref{bdc2} for vacuum solutions of the ${\cal N} = \frac{1}{2}$ model with asymptotically constant masses, $\mu_{Li}$ and $\mu_{Ri}$, becomes 
\begin{align}\label{bdc3}
\mu_{L1} \mu_{L2} \mu_{L3} \,\sum_{n=1}^\infty n(n^2 -1) N^{(L)}_n = \mu_{R1} \mu_{R2} \mu_{R3} \,\sum_{n=1}^\infty n(n^2 -1) N^{(R)}_n . 
\end{align}
Since the  constant parameters, $\mu_{Li}$ and $\mu_{Ri}$, are in general different,  the occupation numbers, $\{N_n^{(L)} \}$ and $\{N_n^{(R)} \} $, can be different. This   implies that some vacuum solutions of the equation \eqref{vacEq-2}, which satisfy the boundary condition \eqref{bdc3}, can interpolate between two different supersymmetric vacua of the ${\cal N} = 1^*$ mSYM theory. The boundary condition \eqref{bdc3} is similar with that in the ${\cal N} = 3$ ImABJM  
model~\cite{Kim:2019kns}.  

 As we have mentioned in the introduction, the homogeneous mass-deformations at the asymptotic boundaries arise from magnetic RR 3-form flux in the dual type IIB supergravity. In \cite{Bena:2015fev}, it was verified that this 3-form flux is determined by the antisymmetric 3-form tensor $T_{abc}$ depicted in \eqref{threeTen}. The magnetic RR 3-form flux is dual to an electric RR
7-form flux, which couples to multiple D3-branes due to the Myers  effect~\cite{Myers:1999ps}. Each block in the vacuum solution in \eqref{phi456} and \eqref{Ti} is interpreted as the equation for non-commutative two sphere in the 456-direction.  
In \cite{Polchinski:2000uf}, it was shown that the vacuum solution \eqref{phi456} corresponds to adding a set of D5-branes with topology ${\cal M}^4 \times S^2$ in the dual AdS$_5\times S^5$ geometry in type IIB supergravity, where $S^2$ denotes the non-commutative or the fuzzy two-sphere. In the solution \eqref{constr1}  $n_k$ D3-branes are polarized into a D5-brane, where the radius of the fuzzy $S^2$ is proportional to $n_k$ in the large $n_k$ limit. See $e.g.$ \cite{Kim:2011qv} for the details. As a result of the vacuum solution \eqref{phi456}, $N$ D3-branes are polarized into $l$ D5-branes with the fuzzy-two spheres having different radii,
for instance, for the $n_k$-dimensional irreducible representation the radius is given by\begin{align}
R_k^2\sim  \frac{ \mu_{01} \mu_{02} \mu_{03}}{4N} n_k(n_k^2 -1). 
\end{align}  
Therefore, the boundary condition in \eqref{bdc3} is a relation between the radii of the set of the fuzzy-two spheres at the left boundary, which are classified by the occupation numbers $\{N^{(L)}_{n}\}$, and those at the right boundary, which are classified by  $\{N^{(R)}_{n}\}$.  The inhomogeneous vacuum solutions interpolating between the two homogeneous asymptotic solutions, describe the variations of the radii of the fuzzy-two spheres along the spatial coordinate $x$.

\subsection{Vacuum solutions for periodic mass functions}\label{pvs}
In this subsection, we obtain  vacuum solutions for spatially periodic mass-functions, 
 \begin{align}\label{mass-fun}
 \mu_i(x)=m_{i0}+m_i(x)\qquad {\rm with}~\int_0^{\tau}m_i(x)dx=0,
 \end{align}
where $m_{i0}$'s are  positive constant mass parameters and $m_i(x)$'s are periodic functions with a period $\tau$ satisfying the relation $\int_{x_0}^{x_0+\tau} m_i(x) dx = 0$ for an arbitrary position $x_0$. Thus $m_{i0}$'s denote the reference values of the mass functions $m_i(x)$.   Then the boundary condition \eqref{bdc2} is trivially satisfied, since one may take $(x_R - x_L)$ as an integer times the periodicity $\tau$ of the periodic mass functions. So the remaining vacuum equations are the coupled differential equations in \eqref{vacEq-2}.

To solve the vacuum equations by using a similar method as in \cite{Kim:2019kns},  we redefine the complex scalar fields  as 
\begin{align}\label{vacSol-1}
& \Phi_i(x)=e^{K_i(x)}\tilde\Phi_i(x),
 \nonumber\\
&K_i(x)=m_{i0}(\xi_i-x)- \Lambda_i(x), \nonumber\\
&  \Lambda_i = \int^x m_i(x')dx',
 \end{align}
where we introduce the monotonically increasing function $\xi_i$ as coordinates, which satisfies the relation 
\begin{align}\label{dxdx}
\frac{d\xi_i}{dx} = e^{K_i - \sum_{ i' \ne i} K_{ i'}}.
\end{align} 
Inserting \eqref{vacSol-1} into \eqref{vacEq-2}, one can formally write the $i$-th vacuum equation as 
\begin{align}\label{tPhieq}
& e^{\sum_{i'\ne i} K_{i'}}\Big(\frac{d\tilde \Phi_i^\dagger}{d\xi_i} +\frac{1}{2} \sum_{j,k=1}^3 \epsilon_{ijk}[\tilde \Phi_j,\tilde\Phi_k] + m_{i0} \tilde\Phi^\dagger_i \Big)=e^{  \sum_{ i' \ne i} K_{ i'} }\mu_i \big(\tilde \Phi_i^\dagger + \tilde \Phi_i\big), 
\nonumber\\
&\sum_{i=1}^3e^{2 K_i}[\tilde \Phi_i,\, \tilde \Phi_i^\dagger] = 0. 
\end{align} 
We consider the anti-Hermitian configurations satisfying the condition $\tilde\Phi_i^\dagger = -\tilde \Phi_i$, $i.e.$, $\Phi_i^\dagger = -\Phi_i$, which are compatible with the homogeneous vacuum solution in \eqref{Phi123}.   Then the vacuum equation in \eqref{tPhieq} is reduced to  
\begin{align}\label{tPeq2}
\frac{d\tilde \Phi_i}{d\xi_i} -\frac{1}{2} \sum_{j,k=1}^3 \epsilon_{ijk}[\tilde \Phi_j,\tilde\Phi_k] + m_{i0} \tilde\Phi_i=0.
\end{align}

As we see in \eqref{dxdx}, the functions $\xi_i(x)$'s are coupled to each other for different mass functions $\mu_i(x)$'s. In this subsection, we consider a simple case, $\mu(x) = \mu_1 (x) = \mu_2(x) = \mu_3(x)$ and follow the analysis given in \cite{Kim:2019kns}.  
Then the relations in \eqref{vacSol-1} are written as 
\begin{align}\label{vacSol-2}
& \Phi_i(x)=e^{K(x)}\tilde\Phi_i(x),
 \nonumber\\
&K(x)=m_{0}(\xi - x)- \Lambda(x)\qquad  {\rm with }\,\,\Lambda = \int^x m(x')dx',
 \end{align}  
 where $\mu(x) = m_0 + m(x) $ with  $\xi(x)$  satisfies the relation 
\begin{align}\label{dxdx2}
\left(\frac{d\xi}{dx}\right)(x) = e^{-K(x)}= e^{-m_{0}(\xi - x)+ \Lambda(x)}.
\end{align}
From the relation $\Phi_i(x) = \left(\frac{dx}{d\xi}  \right)\tilde\Phi_i(x)$, we notice that since $\Phi_i(x)$'s are periodic for the periodic mass function, so are $\tilde \Phi_i(x)$'s if $\left(\frac{d\xi}{dx} \right)$ is periodic. In order to show the periodic behavior of $\left(\frac{ d\xi}{dx}\right)$, we solve the differential equation \eqref{dxdx2}. Then we obtain 
\begin{align}\label{dxdx3}
\left(\frac{dx}{d\xi}\right)(x) = m_0 \int_{-\infty}^x e^{m_0 (x'- x) + (\Lambda(x') - \Lambda(x))} dx',
\end{align} 
where we selected the integration range to have the relation $\xi\to x$ in the limit $m(x)\to 0$. That is, for the constant mass limit $m(x) \to 0$, the differential equation \eqref{tPeq2} is reduced to the vacuum equation of the ${\cal N} = 1^*$ theory with non-vanishing masses given in \eqref{cmve}. It corresponds to the limit $K(x)\to 0$, $i.e.$, $\xi\to x$. Using the relation $\Lambda (x) = \Lambda (x+\tau)$ for the periodic mass function, we obtain 
\begin{align}\label{dxdx4}
\left(\frac{dx}{d\xi}\right)(x+\tau) &= m_0 \int_{-\infty}^{x+\tau} e^{m_0 (x'- x-\tau) + (\Lambda(x') - \Lambda(x))} dx'
\nonumber\\
&=m_0 \int_{-\infty}^{x} e^{m_0 (x''- x) + (\Lambda(x'') - \Lambda(x))} dx'' = \left(\frac{dx}{d\xi}\right)(x), 
\end{align} 
where we used the change of the integration variable by introducing $ x''= x' - \tau$. Therefore, as we see in \eqref{dxdx4}, $\left(\frac{dx}{d\xi}\right)$ is a periodic function with the periodicity $\tau$.  As we discussed previously,  $\tilde \Phi(x)$'s are also periodic functions.

Except for the first term of the left-hand side in \eqref{tPeq2}, the equations are same with the vacuum equations of the ${\cal N} = 1^*$ theory with non-vanishing constant masses. For the constant mass case with $m_0\ne 0$, we know that the only regular periodic solutions satisfying the \eqref{tPeq2} were given in \eqref{Phi123}. That is, $\frac{d\tilde \Phi_i}{d\xi} = 0$. On the other hand, for $m_0 = 0$ case, the equation \eqref{tPeq2} is reduced to 
\begin{align}\label{tPeq4}
\frac{d\tilde \Phi_i}{d\xi_i} -\frac{1}{2} \sum_{j,k=1}^3 \epsilon_{ijk}[\tilde \Phi_j,\tilde\Phi_k] =0.
\end{align}
Except for the term $\frac{d\tilde \Phi_i}{d\xi_i}$ in \eqref{tPeq4}, this equation is  same with the vacuum equation of the original ${\cal N} = 4$ SYM theory without mass deformation. In the SYM theory, we know that the only regular periodic solutions satisfying the equation \eqref{tPeq4} are constant diagonal matrices, $i.e.$, 
\begin{align}\label{tPD}
\tilde \Phi^i_{D} = {\rm diag} (a^i_1, a^i_2, \cdots, a^i_N ).
\end{align}
Again, $\frac{d\tilde \Phi_i}{d\xi_i}=0$. We show several explicit vacuum solutions for some representative periodic mass functions. 

\noindent
$\bullet$ \underline{$\mu(x) = m_1 \sin q x$ case:}
We consider a mass function 
\begin{align}
\mu(x) = m_1 \sin qx, 
\end{align}
where $m_1>0$ and $q = 2\pi n/(x_R - x_L)$ for some integer $n$. In this case, we cannot use the relation \eqref{dxdx3}, since $m_0=0$. From \eqref{vacSol-2}, we simply obtain 
\begin{align}
\Phi_i(x) = e^{ \frac{m_1}{q} \cos q x} \tilde \Phi^i_{D},
\end{align} 
where $\tilde \Phi^i_{D}$'s are given in \eqref{tPD}. \\

\noindent
$\bullet$ \underline{$\mu(x) = m_0 + m_1 \sin q x$ case:}
In this case, we use the relation 
\begin{align}
\Phi_i(x) = \left(\frac{dx}{d\xi}  \right) \tilde\Phi_i(x),
\end{align}
where we use  \eqref{Phi123} and \eqref{dxdx3}, and then obtain 
\begin{align}
\left(\frac{dx}{d\xi}\right)_x &=m_0 \int_{-\infty}^x e^{m_0(x'-x) - \frac{m_1}{q} \left(\cos q x' - \cos q x\right)} dx', \nonumber\\
\tilde\Phi_i(x) &= -i m_0 T_i.
\end{align}

\section{Conclusion}
In this paper, we obtained 4-dimensional ${\cal N}=\frac12$  and ${\cal N}=1$ ImSYM  models from the ${\cal N}=1^*$ and ${\cal N}=2^*$ mSYM theories, respectively, by allowing the variation of the mass parameters along a single spatial direction $x$. The Killing spinor equations for the ImSYM models lead to the supersymmetric vacuum equations as well as a  boundary condition for the solutions at the two asymptotic boundaries of the coordinate $x$. For constant mass parameters,  the vacuum solutions of the  ${\cal N} = 1^*$ theory with all non-vanishing mass parameters contain only the Higgs branch solutions and they are expressed in terms of $N$-dimensional (in general reducible) representations of the $su(2)$ algebra. Based on this fact, we discussed two types of vacuum solutions of ${\cal N}=\frac12$ ImSYM    model by considering two kinds of mass functions, which are asymptotically constant mass functions and periodic mass functions.  

For asymptotically constant mass functions, the inhomogeneous solutions approach the homogeneous solutions of the ${\cal N}=1^*$ mSYM theory at the two asymptotic boundaries of the spatial coordinate $x$. In general, those two homogeneous solutions can be different, and then the inhomogeneous solutions can interpolate between two different supersymmetric vacua of the ${\cal N} = 1^*$ mSYM theory. We showed that the two distinct vacuum solutions of  the ${\cal N} = 1^*$ mSYM theory at the two boundaries of the spatial coordinate $x$ can be connected to each other by the boundary condition. For these mass functions, we  also gave  possible  brane interpretations for the vacuum solutions. 

In the case of the periodic mass functions, we proposed periodic solutions, which trivially satisfy the asymptotic boundary condition. The periodic solutions are given by $\Phi_i(x)=e^{K_i(x)} \tilde\Phi_i (x),~(i=1,2,3)$, where $e^{K_i(x)}$'s are expected to be periodic functions, and $\tilde\Phi_i$'s with non-vanishing reference masses are the solutions for the ${\cal N}=1^*$ mSYM theory. In this paper, we considered a simpler periodic mass function $\mu (x) \equiv  m_0 + m(x) = \mu_1(x) = \mu_2 (x) = \mu_3 (x)$, and obtained  general vacuum solutions by showing the periodicity of the overall factor $e^{K_i(x)}$. For $m_0 = 0$ case, the vacuum configurations have constant diagonal components  only with an appropriate overall periodic function. On the other hand, for $m_0\ne 0$ case, the vacuum configurations are determined by $\tilde\Phi_i (x)\sim m_0 T_i$ and the overall periodic functions.

We obtained the boundary conditions for asymptotically constant mass functions, without explicit solutions. As we discussed in the subsection \ref{convac}, the boundary conditions can connect different asymptotic vacua in the ${\cal N} = 1^*$ theory with all non-vanishing masses. It would be interesting if we obtain some explicit form of the interpolating vacuum solutions. Corresponding dual gravity solutions can also exist in type IIB supergravity, though finding such solutions would be non-trivial, since those solutions require some information for the transverse space. It means consistent truncation can be very complicated in this case. Applying the gauge/gravity duality established in \cite{Arav:2020obl} to various vacuum solutions, for instance, the periodic vacuum solutions in this paper, is also interesting.

\section*{Acknowledgements}
 
We would like to thank Kyung Kiu Kim  for useful discussions. This work was supported by the National Research Foundation of Korea(NRF) grant with grant number  NRF-2019R1F1A1056815 (Y.K.) and NRF-2019R1F1A1059220, NRF-2020R1A2C1014371,  NRF-2019R1A6A1A10073079 (O.K.). O.K.
would like to thank the APCTP topical research program, `Quantum Geometry and Duality 2020'  for the hospitality during the visit, where part of this
work was done.
 

\appendix
\section{${\cal N} = 1$ ImSYM Model }
As we have mentioned in section 2, the ${\cal N}=2^*$ mSYM theory is obtained from the ${\cal N}=1^*$ theory by setting $\mu_1 = \mu_2=\mu$, and   $\mu_3 = 0$. The inhomogeneous extension of the ${\cal N}=2^*$ theory means that the mass parameter is space-dependent. Then the supersymmetry variation of the mass-deformed Lagrangian $({\cal  L_{\rm SYM}}+{\cal L}_\mu)$ is given by
\begin{align}\label{N=2susy-b}
(\delta_\epsilon+\delta'_\epsilon)({\cal  L_{\rm SYM}}+{\cal L}_\mu)&=2i(\partial_\alpha\mu){\rm tr}\Big[\Big(-\sum_{a=1}^{3}+\sum_{a=4}^{6}\Big)\phi_a\bar\psi_m \big(\Gamma_a^{mi}P_{+}
 +\bar\Gamma_a^{mi}P_{-}\big)\gamma^\alpha\epsilon_i\Big],
\end{align} where $m,\cdots=\{1,2\}$ and  $i,\cdots=\{3,4\}$.
From \eqref{Gamma} we see that $\Gamma_a^{ij}=0$ for $a=1,3,4,6$ and $\Gamma_a^{mi}=0$ for $a=2,5$. Therefore, one can write
\begin{align}
(\delta_\epsilon+\delta'_\epsilon)({\cal  L_{\rm SYM}}+{\cal L}_\mu)&=2i(\partial_\alpha \mu){\rm tr}\Big[\Big(-\sum_{a=1,3} +\sum_{a=4,6}\Big)\phi_a\bar\psi_p \big( \Gamma_a^{pi}P_{+}
 +\bar\Gamma_a^{pi}P_{-} \big)\gamma^\alpha \epsilon_i \Big] 
\nn\\
  & ={\rm tr} \big[-2i \partial_\alpha J_{ab}\phi_a\bar\psi_p \big(\Gamma_b^{pi}P_{+}
 +\bar\Gamma_b^{pi}P_{-}\big)\gamma^\alpha \epsilon_i \big],
\end{align}
where $J_{ab}={\rm diag}(\mu,0,\mu,-\mu,0,-\mu)$.

Assuming the mass parameter is inhomogeneous only along one space direction $\mu=\mu(x)$, and using the projection condition $\gamma^1\epsilon_i=\epsilon_i$, which breaks half the supersymmetry, we obtain 
\begin{align}
&(\delta_\epsilon+\delta'_\epsilon)\big({\cal  L_{\rm SYM}}+{\cal L}_\mu\big)={\rm tr}\Big[ 2J'_{ab}\phi_a\Big(-i\bar\psi_p \big(\Gamma_b^{pi}P_{+}
 +\bar\Gamma_b^{pi}P_{-}\big)\epsilon_i\Big)\Big] ={\rm tr}\big(2 J'_{ab}\phi_a\delta\phi_b\big).
\end{align}
This term is cancelled by introducing the following additional  mass Lagrangian term for the scalar field 
\begin{align}
{\cal L}_J=-{\rm tr }\big( J'_{ab}\phi_a\phi_b\big).
\end{align}
The resulting theory is the ${\cal N}=1$ ImSYM model. 

Following the same procedure as in the case of the ${\cal N}=1^*$ theory in section 2, we obtain the vacuum equations and the boundary condition for the ${\cal N} =1$ ImSYM model,
 \begin{align}\label{vacN2}
 &\phi_a'\bar\Gamma_a^{pi}-\frac{ig}2[\phi_a,
 \phi_b]\bar\Gamma^{pr}_{a}\Gamma^{ri}_{b}+\mu_{pr}\phi_a\Gamma_a^{ri}=0,\qquad \big( J_{ab}\phi_a\phi_b+\frac{ig}{3}\tilde T_{abc}\phi_a[\phi_b,\phi_c]\big)\big|_{\rm boundary}=0,
 \end{align}
 where $\tilde T_{135}=-\tilde T_{456}=-1$ are the only non-zero elements of the constant antisymmetric tensor $\tilde T_{abc}$. Inserting $p=\{1,\cdots,4\}$ and $i=\{3,4\}$ and separating the real and imaginary parts of the first equation in \eqref{vacN2}, we obtain 
 
 \begin{align}
& \phi_3'-ig[\phi_1,\phi_5]-\mu\phi_3=0,\qquad\qquad [\phi_1,\phi_2]=0, \nn \\
&\phi_6'-ig[\phi_4,\phi_5]+\mu\phi_6=0,\qquad\qquad [\phi_2,\phi_4]=0,\nn \\
 & \phi_4'-ig[\phi_5,\phi_6]+\mu\phi_4=0,\qquad\qquad [\phi_2,\phi_6]=0,\nn \\
 &\phi_1'+ig[\phi_3,\phi_5]-\mu\phi_1=0,\qquad\qquad [\phi_2,\phi_3]=0,\nn \\
 & [\phi_1,\phi_4]-[\phi_3,\phi_6]=0, ~~\qquad\quad\qquad[\phi_2,\phi_5]=0, \nn \\
& \phi_5'+ig\big([\phi_1,\phi_3]+[\phi_4,\phi_6] \big)=0,~~\quad\phi_2'=0,\nn \\
 &[\phi_1,\phi_6]+[\phi_3,\phi_4]=0.
 \end{align}
 In the absence of the inhomogeneity, the solutions to these vacuum equations contain the Higgs branch solutions for the massive scalar fields $(\phi_1,\phi_3,\phi_4,\phi_6)$ and the Coulomb branch solutions for the massless ones $(\phi_2,\phi_3)$.

\end{document}